\documentclass[11pt,aps,showpacs,nofootinbib,prd,aps,epsf,floats,
               amsmath,amssymb,amsfonts]{revtex4-1}
\usepackage{amsmath, amssymb}
\usepackage{slashbox}

\newcommand{\mathsym}[1]{{}}

\usepackage{graphicx}
\usepackage{amsmath}
\usepackage{amssymb}
\usepackage{color}
\usepackage{bm}
\newcommand{\be}{\begin{equation}}
\newcommand{\ee}{\end{equation}}
\newcommand{\bea}{\begin{eqnarray}}
\newcommand{\eea}{\end{eqnarray}}

\newcommand{\rem}[1]{}
\newsavebox{\PSLASH}
 \sbox{\PSLASH}{$p$\hspace{-1.8mm}/}
 
\renewcommand{\theequation}{\thesection.\arabic{equation}}
\newcounter{saveeqn}
\newcommand{\add}{\addtocounter{equation}{1}}
\newcommand{\alpheqn}{\setcounter{saveeqn}{\value{equation}}%
\setcounter{equation}{0}%
\renewcommand{\theequation}{\mbox{\thesection.\arabic{saveeqn}{\alph{equation}}}}}
\newcommand{\reseteqn}{\setcounter{equation}{\value{saveeqn}}%
\renewcommand{\theequation}{\thesection.\arabic{equation}}}

 \newsavebox{\notrightarrow}
 \sbox{\notrightarrow}{$\to$\hspace{-4mm}/}
 
 \newsavebox{\PARTIALSLASH}
 \sbox{\PARTIALSLASH}{$\partial$\hspace{-1.6mm}/}
 
 \newsavebox{\ASLASH}
 \sbox{\ASLASH}{$A$\hspace{-2.1mm}/}
 
 \newsavebox{\KSLASH}
 \sbox{\KSLASH}{$k$\hspace{-1.8mm}/}
 
 \newsavebox{\LSLASH}
 \sbox{\LSLASH}{$\ell$\hspace{-1.8mm}/}
 
 \newsavebox{\QSLASH}
 \sbox{\QSLASH}{$q$\hspace{-1.8mm}/}
 
 \newsavebox{\DSLASH}
 \sbox{\DSLASH}{$D$\hspace{-2.2mm}/}
 
 \newsavebox{\DbfSLASH}
 \sbox{\DbfSLASH}{${\mathbf D}$\hspace{-2.8mm}/}
 
 \newsavebox{\DELVECRIGHT}
 \sbox{\DELVECRIGHT}{$\stackrel{\rightarrow}{\partial}$}
 
 \newcommand{\blue}{\IfColor{\textCadetBlue}{}}
\newcommand{\black}{\IfColor{\textBlack}{}}
\newcommand{\red}{\IfColor{\textRed}{}}
\newcommand{\green}{\IfColor{\textOliveGreen}{}}
\newcommand{\lila}{\IfColor{\textRedViolet}{}}

\begin{document}
\title{Exploring anomalous $HZ\gamma $ couplings in $\gamma $-proton collisions at the LHC}

\author{S. Taheri Monfared ${}^{a,b}$}\email{sara.taheri@ipm.ir}
\author{Sh. Fayazbakhsh ${}^{a,b}$}\email{shfayazbakhsh@ipm.ir}
\author{M. Mohammadi Najafabadi ${}^{a}$}\email{mojtaba@ipm.ir}

\affiliation{${}^a$School of Particles and Accelerators, Institute for Research in Fundamental Sciences (IPM), P.O. Box 19395-5531, Tehran-Iran}

\affiliation{${}^b$Department of Physics, Faculty of Basic Science, Islamic Azad University Central Tehran Branch
(IAUCTB), P.O. Box 14676-86831, Tehran, Iran}

\begin{abstract}
The $HZ\gamma $ coupling, which is highly sensitive to the new physics beyond the standard model, is studied through the process $pp\rightarrow p\gamma p\rightarrow pHX$ at the LHC. To this purpose, an effective Lagrangian, in a model independent approach, with dimension six operators is considered in this paper. New interaction terms regarding beyond the standard model physics include the Higgs boson anomalous vertices in both CP-even and CP-odd structures. A detailed numerical analysis is performed to scrutinize the accurate constraints on the effective $HZ\gamma $ couplings and to discuss how far the corresponding bounds can be improved. This is achieved by testing all the efficient Higgs decay channels and increasing the integrated luminosity at three different forward detector acceptance regions. The numerical results propose that the Higgs photoproduction at the LHC, as a complementary channel, has a great potential of exploring the $HZ\gamma $ couplings.
\end{abstract}
\maketitle

\section{Introduction}\label{sec1}
The standard model (SM) remarkable predictions are currently approved to elucidate several experimental phenomena in particle physics at low energies. However, there is a variety of physical points which cannot be explained by this effective theory and this is a sensible reason to go beyond the SM (BSM) \cite{Masso:2014xra,Belanger:2015nma}. Although there has not been observed any direct evidence of new physics (NP) at the LHC run-I, it is anticipated to discover signals of NP at the LHC run-II with the help of new observables \cite{Belusca-Maito:2015lna}.
\par
According to the matter content of the SM and the known interaction terms, a number of frameworks are classified to probe likely NP effects at available energies. As one of the current methods, the model independent approach is extensively applicable in such studies. Here, based on the SM symmetry pattern, the conservation of lepton and baryon numbers, and the spontaneous electroweak symmetry breaking (EWSB) in the Higgs mechanism, an effective Lagrangian is formed from NP interactions between the elementary particles \cite{Buchalla:2015wfa}. Indeed, integrating out heavy degrees of  freedom at the BSM scale, $\Lambda $, some residual interaction terms are obtained including the gauge invariant non-renormalizable effective operators. Among these NP operators, the Higgs boson anomalous interactions are also theoretically studied in the literature \cite{Chen:2014ona,Falkowski:2015fla,Englert:2015hrx,Hankele:2006ma,Masso:2012eq,Han:2005pu,Eboli:1999pt,Zeppenfeld:2002ng,Zhang:2003it,Cao:2015fra,Senol:2012fc,Cakir:2013bxa,Heinemeyer:2010gs}.
\par
Following the discovery of the Higgs boson at the LHC \cite{Chatrchyan:2012xdj,Aad:2012tfa,Aad:2015zhl}, describing the properties of this particle is crucial to characterize the nature of the EWSB and to explore possible BSM physics. In the SM framework, the Higgs boson, the massless photon, and the $Z$ boson couple indirectly via loop diagrams, containing massive charged particles. The SM prediction for the decay width of the Higgs particle in the $H\rightarrow Z\gamma $ channel is given by
%
\begin{eqnarray}
\label{width}
\Gamma (H\rightarrow Z\gamma )=\frac{m_{H}^{3}}{16
\pi }\big(1-\frac{m_{Z}^{2}}{m_{H}^{2}}
\big)^{3}\times |G_{\text{SM}}|^{2},
\end{eqnarray}
where, $m_{H}$ and $m_{Z}$ are the masses of the Higgs and $Z$ boson fields, respectively \cite{Masso:2012eq,Djouadi:1996yq}. $G_{\text{SM}}$ includes the $W$ boson and top quark loops contributions and it amounts to around $G_{\text{SM}}\simeq -4.1\times 10^{-5}~\mbox{GeV}^{-1}$. The width in Eq. (\ref{width}) is almost equal to $6\times 10^{-6}~\mbox{GeV}$, which is corresponding to a branching fraction, $Br(H\rightarrow Z\gamma )=1.55 \times 10^{-3}$, at $m_{H}=125~\mbox{GeV}$. The CMS \cite{Chatrchyan:2013vaa} (ATLAS \cite{Aad:2014fia}) collaboration has reported that the observed $95\%$ confidence level (C.L.) decay width for the process $H\rightarrow Z\gamma $ is 10 (11) times more than the value predicted by the SM. Therefore, the rare $HZ\gamma $ vertex is highly sensitive to NP effects from beyond TeV scale \cite{Bergstrom:1985hp,Cahn:1978nz,Ellis:1975ap,Djouadi:1996yq,Gehrmann:2015dua}. Moreover, the $HZ\gamma $ coupling allows one to consider different kinds of NP hypotheses. Some authors suggest that different particles may circulate in the loop diagrams \cite{Chen:2013vi,Chiang:2012qz,Carena:2012xa} and the Higgs boson is described as a non-SM scalar field \cite{Low:2011gn,Low:2012rj} or a massive composite state \cite{Azatov:2013ura}.
\par
A lot of analyses, commonly performed to explore the NP effects, include CP-even effective operators \cite{Corbett:2012ja,Banerjee:2013apa,GonzalezGarcia:1999fq}. However, there are many lines of evidence indicating the CP violation in weak interactions as well as in astronomical observables which are not completely predicted by the Kobayashi--Maskawa theoretical mechanism. Hence, the existence of large amount of CP-violating interactions coming from NP effects is remarkable, especially in order to explain the baryon asymmetry in the universe \cite{Sakharov:1967dj,Riotto:1998bt}. Another motivating aspect is that CP-even and CP-odd anomalous $HZ\gamma $ couplings are all related to higher-dimension NP operators and have also the same order of magnitude \cite{Buchmuller:1985jz,Li:2015kxc,Dwivedi:2015nta,Inoue:2014nva,Chen:2015gaa}. The constraints on the anomalous gauge-Higgs couplings and their collider implications have been widely studied in the literature either with CP-even \cite{Masso:2012eq,Han:2005pu,Zhang:2003it} and CP-odd \cite{Dwivedi:2015nta} dimension six operators.
\par
In this paper, we concentrate on extracting sensitivity of the Higgs production cross section to the anomalous $HZ\gamma $ vertex in single diffractive interactions at the LHC. Here, one of the protons in a $pp$ collision dissociates while the other one remains intact and scatters at small angles. The latter loses a fractional proton energy, $\xi $. The parameter $\xi $ specifies the detector acceptance region in which forward intact protons are observed. Indeed, $\xi $ is determined by the difference between the momentum of the incoming proton, $p$, and that of the intact scattered one, $p'$, i.e., $\xi =(|\vec{p}|-|\vec{p}'|)/|\vec{p}|$. At the LHC energy scale, to a good approximation, the equality $\xi =E_{\gamma }/E_{p}$ arises, where $E_{p}$ and $E_{\gamma }$ are the energies of the incoming proton and the emitted quasireal photon, respectively. Three different classes of the acceptance region according to the CMS and ATLAS scenarios are considered as $0.0015<\xi <0.5$, $0.0015<\xi <0.15$, and $0.1<\xi <0.5$ \cite{Albrow:2008pn,CERN-TOTEM-NOTE}. Recently, the NP effects in the diffractive interactions are discussed in Refs. \cite{Fayazbakhsh:2015xba,Tasevsky:2013iea,Tasevsky:2014cpa}. In what follows, the $HZ\gamma $ coupling is studied through the process $pp\rightarrow p\gamma p\rightarrow pHX$ at the LHC using the effective Lagrangian approach. Both the CP-conserving and -violating interactions arising from dimension six operators are considered for three detector acceptance regions at center of mass energies $\sqrt{s}=14, 100~\mbox{TeV}$.
\par
The present paper is organized as follows: In Sec.~\ref{sec2}, we will introduce the effective Lagrangian which includes anomalous interactions in the Higgs sector with $HZ\gamma $ couplings. The cross section of the collision $pp\rightarrow p\gamma p\rightarrow pHX$ at the LHC with the center of mass energies $\sqrt{s}=14, 100~\mbox{TeV}$ in terms of the anomalous couplings are presented in Sec.~\ref{sec3}. The numerical analysis and some estimations of the cross section sensitivity to the Higgs couplings are reported in Sec.~\ref{sec4}. We will determine the constraints expected at the LHC Run-II for the proposed anomalous operators. Sec.~\ref{sec5} is devoted to a discussion on our concluding results.

\section{The effective Lagrangian and anomalous interactions}\label{sec2}
The SM predictions for the $HZ\gamma $ coupling is based on the heavy quarks and $W$ boson loops computations which depend on the masses of circulating particles. To investigate NP additional contributions to the $HZ\gamma $ vertex, we start with an effective Lagrangian involving the effects of non-SM fields interactions. This Lagrangian can be obtained by the generalization of the SM interaction terms, from all dimension four operators to higher-dimension ones. Ignoring possible dimension five operators, which relate to the non-conservation of lepton number \cite{Weinberg:1979sa,Belusca-Maito:2015lna}, the expansion of the effective Lagrangian can be truncated at dimension six operators as follows:
%
\begin{eqnarray}
\mathcal{L}_{\text{eff.}}=\mathcal{L}_{\text{SM}}+\sum
_{i} \frac{c_{i}^{(6)}{\mathcal{O}}_{i}^{(6)}}{
\Lambda^{2}} + H.c.,
\end{eqnarray}
where, $c^{(6)}_{i}$ and $\mathcal{O}^{(6)}_{i}$ represent dimensionless Wilson coefficients and gauge invariant local operators, respectively.
\par
Probing the Higgs properties, we note that $\mathcal{O}^{(6)}_{i}$ include the Higgs anomalous interactions with gauge bosons and fermion fields \cite{Masso:2012eq}. With a scalar doublet, $\Phi $, which will be replaced by the Higgs field, there are seven dimension six relevant operators. Four of these operators have CP-even structures as $\mathcal{{O}}_{VV}=\Phi^{+}V_{\mu \nu }V^{\mu \nu }\Phi $ and $\mathcal{{O}}_{V}=(D_{\mu }\Phi )^{+}V^{\mu \nu }D_{\nu }\Phi $ for $V\equiv B,W$ and the remaining three CP-odd operators are $\mathcal{{O}}_{\tilde{V}V}=\Phi^{+}\tilde{V}_{\mu \nu }V^{\mu \nu }\Phi $ for $V\equiv B,W$ and $\mathcal{{O}}_{\tilde{B}}$, in the notation of Ref.~\cite{Hankele:2006ma}. The covariant derivative and the field strength tensors of gauge fields are $D_{\mu }= \partial_{\mu }+\frac{i}{2}g'B_{\mu }+\frac{i}{2}g\sigma^{i}W^{i}_{ \mu }$ and $B_{\mu \nu }+W_{\mu \nu }=[D_{\mu },D_{\nu }]$, respectively. $\tilde{V}_{\mu \nu }=\frac{1}{2} \varepsilon_{\mu \nu \rho \sigma }V^{\rho \sigma }$ and $ \varepsilon_{\mu \nu \rho \sigma }$ is a totally antisymmetric tensor with $\varepsilon_{0123}=1$. In the following, the operator $\mathcal{{O}}_{BW}$ is disregarded since it has already been stringently constrained by precision electroweak data and the measurements of the triple gauge boson couplings. Very similar to the SM predictions, the accurate bounds on the coefficient of $\mathcal{{O}}_{BW}$ depend on the masses of the Higgs particle and the top quark so the LHC cannot provide more information on this operator \cite{Hagiwara:1993ck}.
\par
After the EWSB, the effective Lagrangian in the Higgs sector, up to the first power of the Higgs boson, is described in terms of the physical fields interactions. In this paper, we study the $HZ\gamma $ vertices and the corresponding interaction terms from the summation of seven aforementioned operators, after the transformation of $B,W$ fields to $A,Z$ bosons, are given by \cite{Hankele:2006ma,Hagiwara:1993ck,Chen:2014ona}
\begin{eqnarray}
\label{eq:LHVV}
{\mathcal{L}}_{\text{eff.}}^{(6)}& =& g_{HZ\gamma
}^{(1)}\partial_{\nu }HZ_{\mu }A^{\mu \nu }+
 g_{HZ\gamma}^{\mbox{{(2)}}} H A_{\mu
\nu } Z^{\mu \nu }+ \tilde{g}_{HZ\gamma} H
\tilde{Z}_{\mu \nu } A^{\mu \nu }+ H.c.,
\end{eqnarray}
where, $g_{HZ\gamma}^{\mbox{{(i)}}}$, $i=1,2$ are the coefficients of the CP-even operators and $\tilde{g}_{HZ\gamma}$ is the coupling regarding the CP-odd interaction term. The new couplings $g_{HZ\gamma}^{\mbox{{(i)}}}$, $i=1,2$ and $\tilde{g}_{HZ\gamma}$ are functions of $m_{W}$, $m_{Z}$, the $SU(2)_{L}$ coupling constant, $g$, the weak mixing angle, $\theta_{W}$, and some dimensionless parameters that should be constrained in searching for NP effects \cite{Achard:2004kn}. However, in a more common way, one may study the constraints on dimensionful coefficients, $g_{HZ\gamma}^{(i)},\tilde{g}_{HZ\gamma}$, $i=1,2$, or some combinations of them instead of the anomalous dimensionless couplings.
\par
If we rewrite the Eq. (\ref{eq:LHVV}) in a compact form such as $\mathcal{L}_{\text{eff.}}^{(6)}=HZ_{\mu }T^{ \mu \nu }A_{\nu }+H.c.$, the $T^{\mu \nu }$ vertex of the $HZ\gamma $ interaction in momentum space is
%
\begin{eqnarray}
\label{eq:vertex}
T^{\mu \nu }(k,Q)&=&\hat{\alpha }(k,Q)Q^{2}g^{\mu \nu }+\alpha_{1}(k,Q)
 [Q\cdot k g^{\mu \nu }-Q^{\mu }k^{\nu }]+\alpha_{2}(k,Q)
\varepsilon^{\mu \nu \rho \sigma }Q_{\rho }k_{\sigma }.
\end{eqnarray}
Here, $k$ and $Q$ denote the $Z$ boson and photon momenta. By plugging the above vertex into Eq. (\ref{eq:LHVV}), the relations $ \hat{\alpha }\equiv -g_{HZ\gamma}^{(1)}$, $\alpha_{1}\equiv -g_{HZ\gamma}^{(1)}+2g_{HZ\gamma}^{(2)}$, and $\alpha_{2}\equiv 2\tilde{g}_{HZ\gamma}$ arise. Practically, $(\hat{\alpha }, \alpha_{1},\alpha_{2})$ are dimensionful independent coefficients whose sizable values would represent NP effects. The nonzero values are possibly induced by heavy particles loops and can change Higgs production cross sections in comparison with the leading order results of the SM, i.e., $\hat{\alpha }^{\text{SM}}=\alpha^{\text{SM}}_{1}=\alpha^{\text{SM}}_{2}=0$
\cite{Hankele:2006ma}.

\section{Higgs production cross sections}\label{sec3}
\begin{figure}[h]
\centerline{
\includegraphics[clip,width=0.45\textwidth]{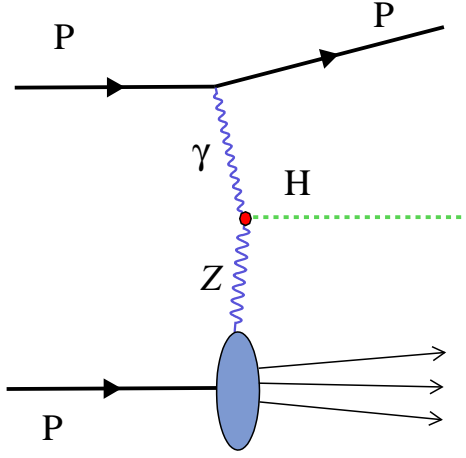}}
\caption{A schematic Feynman diagram of the process $pp\rightarrow pHX$.}
\label{ffeynman}
\end{figure}

\begin{figure}[h]
\centerline{
\includegraphics[clip,width=0.45\textwidth]{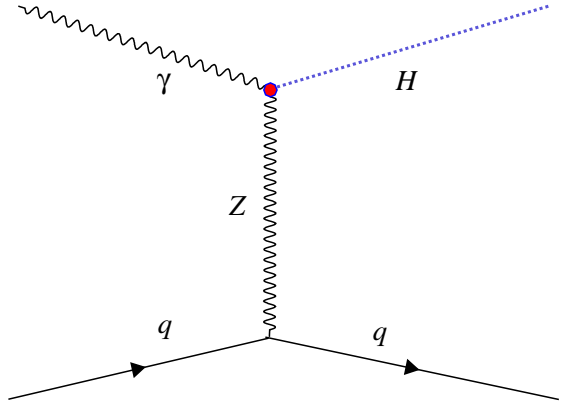}}
\caption{A representative leading order Feynman diagram of the subprocess $\gamma q\rightarrow\gamma Zq\rightarrow Hq$.}
\label{fffeynman}
\end{figure}

The effective Lagrangian introduced in Eq. (\ref{eq:LHVV}) allows the production of a Higgs boson through the subprocess $\gamma q\rightarrow \gamma Zq\rightarrow Hq$ in the collision $pp\rightarrow p\gamma p \rightarrow pHX$. The Feynman diagrams for the main process $pp\rightarrow pHX$ and the subprocess $\gamma q\rightarrow Hq$ at leading order are depicted in Figs. \ref{ffeynman} and \ref{fffeynman}, respectively. The total scattering amplitude, $\overline{|M|}^{2}$ in the quasireal photon approximation, with zero mass photons, is dependent on two of the anomalous couplings, i.e., $\overline{|M|}^{2}=\overline{|M_{1}|}^{2}\alpha_{1}^{2}(k,Q)+\overline{|M_{2}|}^{2}\alpha_{2}^{2}(k,Q)$. The functions $\overline{|M_{1}|}^{2}$ and $\overline{|M_{2}|}^{2}$ with the redefinition of vector and axial-vector couplings, $C^{\pm }_{q}=C_{q,V}^{2}\pm C_{q,A}^{2}$, and using the relation $\hat{s}+\hat{t}+\hat{u}=m_{H}^{2}+2m_{q}^{2}$ read 
%
\begin{eqnarray}
\overline{|M_{1}|}^{2}&=&-\frac{g_{Z}^{2}}{8(m_{Z}^{2}-\hat{t})^{2}}\bigg \{C^{+}_{q}
\hat{t}\bigg((\hat{t}+\hat{s}
-m_{q}^{2}-m_{H}^{2})^{2}+(\hat{s}-m_{q}^{2})^{2}
\bigg)+2C^{-}_{q}m_{q}^{2}(m_{H}^{2}-\hat{t})^{2}\bigg \},
\nonumber
\\
\overline{|M_{2}|}^{2}&=&-\frac{g_{Z}^{2}}{4(m_{Z}^{2}-\hat{t})^{2}}(m_{H}^{2}-\hat{t})^{2}
\bigg \{C^{+}_{q}(\hat{t}-2m_{q}^{2})+4C
^{-}_{q}m_{q}^{2}\bigg \}.
\end{eqnarray}
\par
A technical tool to perform the numerical calculations of a photon-induced subprocess, is the equivalent photon approximation (EPA) method. This is applied to the collisions in the forward direction, where the $Q^{2}/E_{\gamma }^{2}\ll 1$ estimation between the energy, $E_{\gamma }$, and virtuality, $Q^{2}$, of the photon is justified. Hence, the emitted photon is considered as a quasireal particle (see \cite{Baur:2001jj,Piotrzkowski:2000rx,Budnev:1974de} for reviews on the EPA method) whose spectrum is given by
%
\begin{eqnarray}
\label{eq1}
f(E_{\gamma },Q^{2})&=&\frac{dN}{dE_{\gamma }dQ^{2}}
=\frac{
\alpha_{e}}{\pi }\frac{1}{E_{\gamma }Q^{2}}\bigg[
\big(1-\frac{E_{\gamma }}{E_{p}}\big)\big(1-\frac{Q
^{2}_{\text{{min}}}}{Q^{2}}\big)F_{E}+\frac{E
^{2}_{\gamma }}{2E_{p}^{2}}F_{M}\bigg],
\end{eqnarray}
where,
%
\begin{align}
\label{eq2}
&F_{E}=\frac{4m^{2}_{p}G^{2}_{E}+Q^{2}G^{2}_{M}}{4m^{2}_{p}+Q^{2}},
\qquad
F_{M}=G^{2}_{M},
\nonumber
\\
&G^{2}_{E}=\frac{G^{2}_{M}}{\mu^{2}
_{p}}=\big(1+\frac{Q^{2}}{Q_{0}^{2}}\big)^{-4},
\qquad
Q_{0}^{2}=0.71~\mbox{GeV}^{2},
\nonumber
\\
&Q^{2}_{\text{{min}}}=\frac{E_{\gamma }^{2}m_{p}
^{2}}{E_{p}(E_{p}-E_{\gamma })},
\qquad
Q^{2}_{\text{{max}}}=2~\mbox{GeV}^{2}.
\end{align}
Here, $m_{p}$ is the proton mass and $\alpha_{E}$ is the QED fine structure constant \cite{Budnev:1974de,Baur:2001jj,Piotrzkowski:2000rx}. The $F_{E}$ and $F_{M}$ functions are determined by the proton electric and magnetic form factors, respectively. The proton magnetic moment is fixed with the value $\mu^{2}_{p}=7.78$ and from now on in this paper, we impose the relation $E_{\gamma }=E_{p}\xi $ in the EPA. As mentioned above, all the terms of scattering amplitudes proportional to powers of $\hat{\alpha }$ will be eliminated due to the EPA method for the quasireal photons, i.e., $Q^{2}=0$.
\par
For parton distribution functions (PDFs) to generate hard scattering matrix elements, we take the leading order results of three main PDF fitting collaborations, NNPDF3.0 \cite{Ball:2014uwa}, CTEQ14 \cite{Dulat:2015mca}, and MMHT14 \cite{Harland-Lang:2014zoa}, which have provided updates for their global analyses. These PDF sets are precisely compared in Refs. \cite{Rojo:2015acz,Ball:2015oha} and consequently an improved agreement with the former releases is demonstrated. The uncertainty due to the choice of a particular PDF set arises from limited knowledge of the proton structure. It is estimated by performing all computations of the signal cross sections for different PDF sets. According to the PDF4LHC recommendations \cite{Butterworth:2015oua}, uncertainties $0.022\%$, $0.019\%$, and $0.161\%$ are found for the first, second, and third acceptance regions at $\sqrt{s}=14~\mbox{TeV}$, respectively.
\par
\begin{figure}[h]
\centerline{
\includegraphics[clip,width=0.6\textwidth]{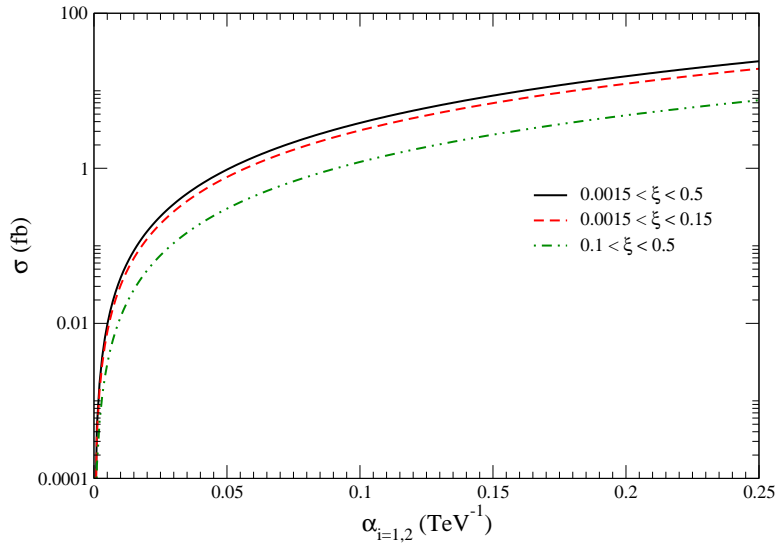}}
\caption{The total cross section of the process $pp\rightarrow pHX$ as a function of the anomalous coupling $\alpha_{1(2)}$ at $\alpha_{2(1)}=0$ and at the center of mass energy $\sqrt{s}=14$ TeV. The curves show the sensitivity for three different acceptance regions remarked on the figure.}
\label{xsection}
\end{figure}

The total cross section is derived by convoluting the subprocess cross section with the photon spectrum in the EPA method and PDF sets as follows:
%
\begin{eqnarray}
\label{cross}
\sigma &=&\sum\limits_{q=u,d,s,c,b}\,
\int_{\omega_{\text{{min}}}}^{\omega_{\text{{max}}}}\frac{
\omega }{2E_{p}y}
d\omega \int_{y_{\text{min}}}^{y_{\text{max}}}dy
\int_{Q^{2}_{1,\text{min}}}^{Q^{2}_{\text{max}}}
dQ_{1}^{2}\, f_{\gamma }\big(y,Q
_{1}^{2}\big)\times f_{q}\big(\frac{\omega^{2}}{4E_{p}y},Q_{2}^{2}\big) \hat{\sigma }_{Z\gamma \to H}(Q
^{2}_{1},\omega ,y),\nonumber\\
\end{eqnarray}
where, the integration limits
%
\begin{align}
\label{intlimits}
&y_{\text{{min}}}=\mbox{Max}\bigg[\frac{
\omega^{2}}{4E_{p}x_{\text{{max}}}},E_{p}\xi_{\text{{min}}}\bigg],
\nonumber
\\
&y_{\text{{max}}}=\mbox{Min}\bigg[\frac{\omega
^{2}}{4E_{p}x_{\text{{min}}}},E_{p}\xi_{\text{{max}}}\bigg],
\nonumber
\\
&\omega_{\text{{min}}}=\mbox{Max}\big[2E_{p}\sqrt{\xi_{\text{{min}}}x_{\text{{min}}}},m_{H}+m_{q}\big],
\nonumber
\\
&\omega_{\text{{max}}}=2E_{p}\sqrt{
\xi_{\text{{max}}}x_{\text{{max}}}},
\end{align}
are imposed. Fig. \ref{xsection} displays the total cross section as a function of the anomalous $HZ\gamma $ couplings at the center of mass energy $\sqrt{s}=14~\mbox{TeV}$. Three separated curves represent the results of three different detector acceptance regions. We found that the functional dependencies of $\sigma $ to the couplings $\alpha_{1}$ and $\alpha_{2}$ are almost similar.
\par
Since the final state includes an intact proton, we have to take into account the effect of survival factor to consider the probability of no additional underlying event activity. The survival factor is important for accurate prediction of the (semi-)exclusive cross section and it depends on the detector performance. We follow the approach in Refs. \cite{Khoze:2000db,Harland-Lang:2016apc}, where it is emphasized that the impact of survival probability sensitively depends on the subprocess, through the specific proton impact parameter dependence. Here the situation can fairly be described as having the evolution component from one proton and the coherent input from the other one. This leads to a $\sim 26\%$ suppression on the expected signal and also background cross section at the scale of the Higgs mass \cite{Harland-Lang:2016apc}.
\par
In the following, the factorization scale, $\mu_{f}$, as well as the renormalization one, $\mu_{r}$, are equal to the Higgs mass, $m_{H}$, which is assumed to be the threshold production scale, i.e., $Q_{2}=\mu_{f}= \mu_{r}=m_{H}$. The uncertainty coming from the factorization/renormalization scale is obtained by doubling, $Q_{2}=2m_{H}$, and halving, $Q_{2}=m_{H}/2$, the threshold scale. Deviations due to the variation of scales are found to be $0.004\%$, $0.012\%$, and $0.054\%$ for the first, second and third regions at $\sqrt{s}=14~\mbox{TeV}$. The third region results contain the largest uncertainty coming from the factorization scale variations and the choice of PDF. Both kind of uncertainties increase with increasing center of mass energy from $\sqrt{s}=14~\mbox{TeV}$ to $\sqrt{s}=100~\mbox{TeV}$ and the uncertainty due to the choice of PDF is larger than the uncertainty arising from the variation of factorization scale.

\section{Sensitivity to the Higgs anomalous couplings}\label{sec4}
In this section, we present the sensitivity of the process $pp\rightarrow p\gamma p\rightarrow pHX$ to the anomalous $HZ\gamma $ couplings for various forward detector acceptance regions and at different values of the integrated luminosity. The final state of the process consists of a Higgs boson, a jet and an intact proton. We study the most important relevant decay channels for the SM Higgs boson: $H\rightarrow \gamma \gamma $, $H\rightarrow W^{+}W^{-}$, $H\rightarrow ZZ$. The SM Higgs boson branching ratios together with corresponding uncertainties can be found in Ref.~\cite{Dittmaier:2012vm}. The SM branching fractions for the Higgs decays to $\gamma \gamma $, $W^{+}W^{-}$, and $ZZ$ are $2.28\times 10^{-3}$, $2.15\times 10^{-1}$, and $2.64\times 10^{-2}$, respectively. The Higgs boson decay to a $b\bar{b}$ pair has the largest branching ratio but it suffers from very large amount of background contributions. Indeed, this decay mode can loosely constrain the anomalous couplings, so we have already excluded this channel in our analysis.
\par
For a given integrated luminosity, $\mathcal{L}_{\mbox{{int}}}$, to assess the sensitivity of the process $pp\rightarrow p\gamma p \rightarrow pHX$ at the LHC, the theoretically predicted number of signal events for each final state, the experimental efficiencies and the expected background events are needed. The number of signal events, $N_{\text{{signal}}}$, reads
%
\begin{eqnarray}
\label{Nsignal}
N_{\text{{signal}}}(\alpha_{1},\alpha_{2}) &=& \sigma (pp \rightarrow
pHX)\times Br(H\rightarrow FF)\times Br(F\rightarrow f_{1}f_{2}...)
\times \mathcal{L}_{\text{{int}}},
\end{eqnarray}
where $F=\gamma ,W^{\pm }$ and $Z$ boson and $f=l^{\pm },\nu_{l}$ (for $F = W,Z$). The values of $Br(F\rightarrow f_{1}f_{2}...)$ in Eq. (\ref{Nsignal}) for the decays of $W$ and $Z$ bosons are $0.05$ and $0.12$, respectively.
\par 
To have a more realistic study, we consider the irreducible photoproduction background ($\gamma +q\rightarrow H+q$) coming from diffractive processes as well as the contribution arising from the reducible photoproduction processes. We found that the reducible photoproduction processes, with different particles in the final state, are expected to be effectively rejected by applying the cuts. Nevertheless, the contribution of the irreducible background is larger than the reducible one after the cuts. The total cross sections of the backgrounds, calculated with CompHEP v4.5.2 package \cite{comphep}, are summarized in Table \ref{tab1}. We perform an explicit calculation of the background subprocesses when one proton is intact and we have $\gamma \gamma +jet$ (for $H\rightarrow \gamma\gamma $ channel), $l_{1}^{\pm }l_{2}^{\mp }\nu_{l_{1}}\nu_{l_{2}}+jets$ (for $H\rightarrow W^{+}W^{-}$ channel), and $l_{1}^{\pm }l_{1}^{\mp }l_{2}^{\pm }l_{2}^{\mp }$ (for $H\rightarrow ZZ$ channel) in the final state.
\begin{table}[h]
	\begin{tabular}{|c||c|c|c|c|c|c|c|}\hline
		&  \multicolumn{3}{c|}{$\sqrt{s}=14$ TeV} &\multicolumn{3}{c|}{$\sqrt{s}=100$ TeV}\\ \hline
		\backslashbox{\ \ \ $\xi$}{Channel} &$H\rightarrow\gamma\gamma$ &$H\rightarrow ZZ$ & $H\rightarrow W^{+}W^{-}$
		&$H\rightarrow\gamma\gamma$ &$H\rightarrow ZZ$ & $H\rightarrow W^{+}W^{-}$   \\ \hline\hline

		$0.0015-0.5$  & 2.5 & 0.5 & 14.4 & 3.4 & 2.1 & 148 \\
		\hline
		$0.0015-0.15$ & 2.7 & 0.45 & 12.6 & 3.6 & 2 & 126 \\
		\hline
		$0.1-0.5$     & 0.1 & 0.09 & 10	& 1.3$\times10^{-4}$ & 0.04 & 24 \\
		\hline
	\end{tabular}
\caption{The total cross sections (unit in fb) of the backgrounds coming from diffractive processes for three final states $\gamma\gamma$, $W^{+}W^{-}$, and $ZZ$ after applying all cuts.}
	\label{tab1}
\end{table}
Now, we obtain the $95\%$ C.L. limits on the Higgs anomalous couplings $\alpha_{1}$ and $\alpha_{2}$ with Poisson statistics at $\sqrt{s}=14,100~\mbox{TeV}$. For a specific integrated luminosity, the expected $95\%$ C.L. upper limits of the number of signal events, $N_{\text{{signal}}}$, is obtained under the assumption that the number of observed events, $N_{\text{{Obs.}}}$, is equal to the number of SM prediction, $N_{\text{{Bkg.}}}$.
\par
The $95\%$ C.L. constraints on $|\alpha_{1}|$, $|\alpha_{2}|$ and the upper limits on the branching ratio of the $H\rightarrow Z\gamma $ decay channel for $\mathcal{L}_{\text{{int}}}=100,300,3000~\mbox{fb}^{-1}$ are presented in Tables \ref{tab2} and \ref{tab3} at $\sqrt{s}=14~\mbox{TeV}$ and $\sqrt{s}=100~\mbox{TeV}$, respectively. The upper limits on the branching ratio corresponding to each coupling constraint are given in parentheses in each column. The bounds corresponding to the decay processes $H\rightarrow \gamma \gamma $, $H\rightarrow W^{+}W^{-}$, $H\rightarrow ZZ$, and the combination of these three Higgs decay channels, are given in separate columns for each detector acceptance region. Here, we do not consider neither the reconstruction nor the acceptance efficiencies.
\par
\begin{figure}[h]
	\centerline{
	\includegraphics[clip,width=0.57\textwidth]{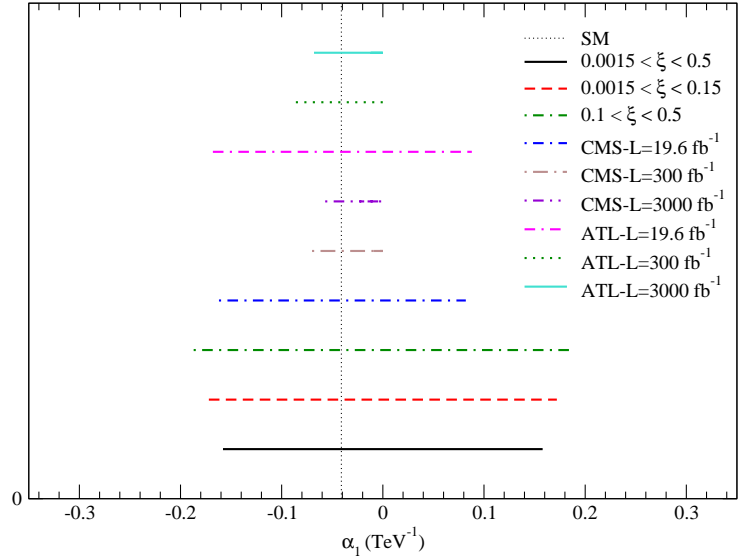}}
\caption{The $95\protect\%$ C.L. constraints on the anomalous $HZ\gamma$ couplings in $H\rightarrow ZZ$ channel at $\sqrt{s}=14$ TeV with an integrated luminosity $\mathcal{L}_{\mbox{\tiny{int}}}=3000$ fb$^{-1}$ for three different acceptance regions at the LHC. The CMS and ATLAS exclusion limits obtained from the Higgs boson rare decay process at $\sqrt{s}=8$ TeV and $\mathcal{L}_{\mbox{\tiny{int}}}=19$ fb$^{-1}$ are shown for comparison. For completeness, the CMS and ATLAS projected allowed regions at $\sqrt{s}=14$ TeV and $\mathcal{L}_{\mbox{\tiny{int}}}=300,3000$ fb$^{-1}$ are also presented.}
\label{Comparison}
\end{figure}

In Fig. \ref{Comparison}, based on the dimension six operator coefficients, the $95\%$ C.L. constraints on the anomalous couplings at $\sqrt{s}=14~\mbox{TeV}$ and for an integrated luminosity $\mathcal{L}_{\text{{int}}}=3000~\mbox{fb}^{-1}$ in $H\rightarrow ZZ$ channel are presented for three different acceptance regions at the LHC. Here, the reconstruction and the acceptance efficiencies are not considered. In the SM, for $m_{H}=125~\mbox{GeV}$ the coupling induced by the $W$ boson and the top quark loops is $\alpha_{1}=G_{\text{SM}}$ \cite{Masso:2012eq}, while the bottom quark contribution is ignored due to its small mass. The CMS (ATLAS) exclusion bound, based on the partial width at $\sqrt{s}=8~\mbox{TeV}$ and $\mathcal{L}_{\text{{int}}}=19.6~\mbox{fb}^{-1}$, is $-0.162\le \alpha_{1}\le 0.082~\mbox{TeV}^{-1}$ ($-0.168\le \alpha_{1}\le 0.088~\mbox{TeV}^{-1}$). Precise measurements on projected performance of upgraded CMS \cite{CMS:2013xfa} (ATLAS \cite{ALTAS:2013}) detectors at the LHC and high luminosity LHC show that the decay process in $H\rightarrow Z\gamma $ channel is expected to be measured at $\sqrt{s}=14~\mbox{TeV}$ with $\sim 62\%$ ($\sim 145 \%$) uncertainties using an integrated luminosity $\mathcal{L}_{\text{{int}}}=300~\mbox{fb}^{-1}$ and $\sim 20 \%$ ($\sim 54 \%$) uncertainties using $\mathcal{L}_{\text{{int}}}=3000~\mbox{fb}^{-1}$ at $95\%$ C.L. Our bounds can also be compared with the ones in Ref.~\cite{Masso:2012eq} in which $|\alpha_{1}|\le 2~\mbox{TeV}^{-1}$ is obtained.
\par
At our proposed channel the sensitivities to probe the $HZ\gamma $ couplings are improved. Reduction strategies for background processes, a realistic analysis with using shape variables, and deriving the background contributions from data would provide more robust results on the exclusion limits of the anomalous couplings.
\begin{table}[h]
	\begin{tabular}{|c||c|c|c|c|c|c|c|c|c|c|}\hline
		
		&  & \multicolumn{4}{c|}{$|\alpha_{1}|=|\alpha_{2}|$\ [TeV$^{-1}$]\ \ $\big(Br(H\to\gamma Z)\big)$} \\ \hline
		$\xi$ & \backslashbox{\ \ \ $\mathcal{L}_{\mbox{\tiny{int}}}$[fb$^{-1}$]}{Channel} &$H\rightarrow \gamma\gamma$ &$H\rightarrow ZZ$ & $H\rightarrow W^{+}W^{-}$ & Combined \\ \hline\hline
		
		& 100  & 0.643 (0.082) & 0.369 (0.033) & 0.450 (0.045) & 0.393 (0.036) \\
		$0.0015-0.5$  & 300  & 0.488 (0.052) & 0.280 (0.021) & 0.342 (0.029) & 0.299 (0.024) \\
		& 3000 & 0.275 (0.021) & 0.158 (0.009) & 0.192 (0.012) & 0.168 (0.010) \\
		\hline
		& 100  & 0.732 (0.102) & 0.402 (0.038) & 0.486 (0.052) & 0.429 (0.042) \\
		$0.0015-0.15$ & 300  & 0.556 (0.064) & 0.305 (0.024) & 0.369 (0.033) & 0.326 (0.027) \\
		& 3000 & 0.313 (0.025) & 0.172 (0.011) & 0.208 (0.014) & 0.183 (0.012) \\
		\hline
		& 100  & 0.531 (0.060) & 0.439 (0.044) & 0.758 (0.108) & 0.635 (0.080) \\
		$0.1-0.5$     & 300  & 0.403 (0.038) & 0.333 (0.028) & 0.576 (0.068) & 0.483 (0.051) \\
		& 3000 & 0.227 (0.016) & 0.187 (0.012) & 0.324 (0.027) & 0.281 (0.022) \\
		\hline
	\end{tabular}	
\caption{The $95\protect\%$ C.L. constraints on the anomalous $HZ\gamma$ couplings, $|\alpha_{1}|$ and $|\alpha_{2}|$, and the upper limits on the branching ratio of the $H\rightarrow Z\gamma$ decay channel in the main process $pp\rightarrow pHX$ at $\sqrt{s}=14$ TeV and for integrated luminosities $\mathcal{L}_{\mbox{\tiny{int}}}=100,300,3000$ fb$^{-1}$. The upper limits on the branching ratio are given in parentheses. The bounds values are presented for three different Higgs decay channels, $H\rightarrow\gamma\gamma$, $H\rightarrow W^{+}W^{-}$, and $H\rightarrow ZZ$ as well as a combined one achievable with $95\%$ C.L. and for three intervals of forward detector acceptance region, $\xi$. The CP-even and CP-odd contributions have the same values. The reconstruction and the acceptance efficiencies are not considered.}
\label{tab2}
\end{table}
\begin{table}[h]
	\begin{tabular}{|c||c|c|c|c|c|c|c|c|c|c|}\hline
		
		&  & \multicolumn{4}{c|}{$|\alpha_{1}|=|\alpha_{2}|$\ [TeV$^{-1}$]\ \ $\big(Br(H\to\gamma Z)\big)$} \\ \hline
		$\xi$ & \backslashbox{\ \ \ $\mathcal{L}_{\mbox{\tiny{int}}}$[fb$^{-1}$]}{Channel} &$H\rightarrow \gamma\gamma$ &$H\rightarrow ZZ$ &$H\rightarrow W^{+}W^{-}$ & Combined \\ \hline\hline
		
		& 100  & 0.173 (0.011) & 0.123 (0.007) & 0.200 (0.013) & 0.169 (0.010)  \\
		$0.0015-0.5$  & 300  & 0.131 (0.008) & 0.093 (0.005) & 0.152 (0.009) & 0.128 (0.007)  \\
		& 3000 & 0.074 (0.004) & 0.052 (0.003) & 0.085 (0.005) & 0.072 (0.004)  \\
		\hline
		& 100  & 0.185 (0.012) & 0.128 (0.007) & 0.203 (0.013) & 0.172 (0.011) \\
		$0.0015-0.15$ & 300  & 0.141 (0.008) & 0.098 (0.006) & 0.155 (0.009) & 0.130 (0.008) \\
		& 3000 & 0.079 (0.005) & 0.055 (0.003) & 0.087 (0.005) & 0.073 (0.004) \\
		\hline
		& 100  & 0.0318 (0.0025) & 0.107 (0.006) & 0.297 (0.023) & 0.248 (0.018) \\
		$0.1-0.5$     & 300  & 0.0241 (0.0023) & 0.081 (0.005) & 0.226 (0.016) & 0.192 (0.012) \\
		& 3000 & 0.0136 (0.0020) & 0.046 (0.003) & 0.127 (0.007) & 0.106 (0.006) \\
		\hline
	\end{tabular}
	\caption{The $95\protect\%$ C.L. constraints on the anomalous $HZ\gamma$ couplings, $|\alpha_{1}|$ and $|\alpha_{2}|$, and the upper limits on the branching ratio of the $H\rightarrow Z\gamma$ decay channel in the main process $pp\rightarrow pHX$ at $\sqrt{s}=100$ TeV and for integrated luminosities $\mathcal{L}_{\mbox{\tiny{int}}}=100,300,3000$ fb$^{-1}$. The upper limits on the branching ratio are given in parentheses. See the caption of Table \ref{tab2} for further details.}
	\label{tab3}
\end{table}
\par
\par
To provide more practical limits, we perform an analysis including detector acceptance, resolution effects and pile-up interactions. The experimental efficiencies for each final state is considered. To reconstruct a specific final state phase space at both $\sqrt{s}=14~\mbox{TeV}$ and $\sqrt{s}=100~\mbox{TeV}$, we take a conservative approach and apply the efficiencies extracted based on Run-I experimental results. We use the following reconstruction efficiencies to study the process $pp\rightarrow pHX$ from Ref.~\cite{Englert:2015hrx}, and the references therein: $\epsilon_{H\rightarrow \gamma \gamma }=0.72$, $\epsilon_{H\rightarrow W^{+}W^{-}}=0.9025$, $\epsilon_{H\rightarrow ZZ}=0.815$. In this study we use the signal in the decay channels of Higgs to $\gamma \gamma $, $W^{+}W^{-}$, and $ZZ$, followed by the leptonic decays of $W$ and $Z$. Each channel has its own background composition and this point finally leads to the extraction of different bounds. The acceptance cuts that select the events are imposed on pseudorapidities, $\eta $, and transverse momenta, $p_{T}$, of the final state particles as:
%
\begin{align}
\label{cuts1}
&\mbox{p}_{T}^{\gamma }\geq 20~
\mbox{GeV},\qquad
\mbox{p}_{T}^{{\text{jet}}}\geq 20~
\mbox{GeV},\qquad
\mbox{p}_{T}^{l}\geq 20~\mbox{GeV},
\nonumber
\\
&|\eta^{\gamma }|< 2.5,
\qquad
|\eta^{{\text{jet}}}| < 2.5,
\qquad
|\eta^{l}| < 2.5.
\end{align}
\par
For further background suppression and the enhancement of signal-to-background ratios, the following cuts are differently applied to each decay channel:
%
\begin{align}
\label{cuts2}
&\!\begin{aligned}
\mbox{For}~\gamma \gamma ~\mbox{channel:}
\quad
&100\mbox{GeV}<M_{{\gamma \gamma }}< 150\mbox{GeV},
\\
&\Delta R(\gamma_{1},\gamma_{2})>0.3,
\end{aligned}
\nonumber
\\
&\!\begin{aligned}
\mbox{For}\ ZZ \ \mbox{channel:}
\quad &70 \mbox{ GeV}<M_{{ll}}< 110 \mbox{
GeV},\\
&100 \mbox{ GeV}<M_{{4l}}< 150 \mbox{ GeV},
\\
&\Delta R(l_{i},l_{j})>0.3,
\end{aligned}
\nonumber
\\
&\!\begin{aligned}[b]
\mbox{For}~W^{+}W^{-}~\mbox{channel:}
\quad&\mbox{No mass cut is applied},\\
&\mbox{MET}>40~\mbox{GeV},
\\
&\Delta R(l^{+},l^{-})>0.3.
\end{aligned}
\end{align}
In relations (\ref{cuts2}), $\Delta R_{ij}=\sqrt{(\eta_{i}-\eta_{j})^{2}+(\phi_{i}-\phi_{j})^{2}}$ and missing transverse energy is denoted by MET. Applying the same cuts on the signal events results in the acceptance efficiencies 0.4, 0.1, and 0.25 for $H\rightarrow \gamma \gamma $, $H\rightarrow W^{+}W^{-}$, and $H\rightarrow ZZ$ channels, respectively.
\par
During each bunch crossing at the LHC, more than a proton--proton interaction can occur which is called a pile-up. Protons within the acceptance of the forward detector from pile-up events can be a source of background to our signal process. In particular, it happens when a pile-up event is placed over a hard non-diffractive process with the same final state as the signal. To estimate the contribution of this type of background, the probability of observing such events in the forward detectors needs to be known. The probability for the measurement of a single proton tagged event in forward detectors depends on the detector-beam center distance and the beam optic. Based on the forward detector specifications and the beam properties, this probability could be at the order of $1\mbox{--}2\%$ \cite{Trzebinski:2015bra}.
\par
The $95\%$ C.L. constraints on $|\alpha_{1}|$, $|\alpha_{2}|$ and the upper limits on the branching ratio of the $H\rightarrow Z\gamma $ decay channel in the decay processes $H\rightarrow \gamma \gamma $, $H\rightarrow W^{+}W^{-}$, $H\rightarrow ZZ$, and the combined channel for $\mathcal{L}_{\text{{int}}}=100,300,3000~\mbox{fb}^{-1}$ are demonstrated in Tables \ref{tab4} and \ref{tab5} at $\sqrt{s}=14~\mbox{TeV}$ and $\sqrt{s}=100~\mbox{TeV}$, respectively. The upper limits on the branching ratio corresponding to each coupling constraint are given in parentheses in each column. Both the reconstruction and the acceptance efficiencies are included into bounds estimations. We have considered $1\%$ probability for observing a single tagged event with the hard non-diffractive process with the same final state as three signal channels $\gamma \gamma $, $WW$, and $ZZ$. We observe minor modifications in the upper limits on the anomalous couplings. For instance, in the first detector acceptance region for $\mathcal{L}_{\text{{int}}}=300~\mbox{fb}^{-1}$ and at $\sqrt{s}=14~\mbox{TeV}$, the upper limits 0.844, 0.593, 1.139, and 0.775 change to 0.864, 0.594, 1.141, and 0.779 in $H\rightarrow \gamma \gamma $, $H\rightarrow ZZ$, $H\rightarrow W^{+}W^{-}$, and combined channels, respectively.
\par
The calculated upper limits on the branching ratio of $H\rightarrow Z\gamma $ decay channel can be compared with the existing bound on branching ratios from the CMS \cite{Chatrchyan:2013vaa} (ATLAS \cite{Aad:2014fia}) collaboration measurements, at $\sqrt{s}=8~\mbox{TeV}$ and $\mathcal{L}_{\text{{int}}}=19.6~\mbox{fb}^{-1}$, which is 0.0064 (0.0068).
\par
As expected, similar exclusion intervals are obtained for $\alpha_{1}$ and $\alpha_{2}$. The $H\rightarrow ZZ$ decay channel provides the more restricted bounds due to having smaller backgrounds. Comparing various $\xi $ ranges, we conclude that the least sensitive region for $W^{+}W^{-}$ and $ZZ$ decay channels is the third acceptance interval, while this region provides the most restricted bounds for $\gamma \gamma $ channel. Using higher integrated luminosities and center of mass energies more stringent limits can be established. A conservative estimation of the most theoretical uncertainties is considered in calculating the limits, while taking into account all systematic uncertainties is beyond the scope of this paper.
\begin{table}[h]
	\begin{tabular}{|c||c|c|c|c|c|c|c|c|c|c|}\hline
		
		&  & \multicolumn{4}{c|}{$|\alpha_{1}|=|\alpha_{2}|$\ [TeV$^{-1}$]\ \ $\big(Br(H\to\gamma Z)\big)$} \\ \hline
		$\xi$ & \backslashbox{\ \ \ $\mathcal{L}_{\mbox{\tiny{int}}}$[fb$^{-1}$]}{Channel} &$H\rightarrow \gamma\gamma$ &$H\rightarrow ZZ$ & $H\rightarrow W^{+}W^{-}$ & Combined \\ \hline\hline
		
		& 100  & 1.187 (0.219) & 0.785 (0.114) & 1.508 (0.310) & 1.035 (0.178) \\
		$0.0015-0.5$  & 300  & 0.864 (0.133) & 0.594 (0.072) & 1.141 (0.206) & 0.779 (0.113) \\
		& 3000 & 0.476 (0.050) & 0.333 (0.028) & 0.640 (0.081) & 0.436 (0.043) \\
		\hline
		& 100  & 1.346 (0.264) & 0.855 (0.131) & 1.632 (0.345) & 1.128 (0.203) \\
		$0.0015-0.15$ & 300  & 0.984 (0.164) & 0.647 (0.083) & 1.234 (0.233) & 0.848 (0.129) \\
		& 3000 & 0.542 (0.061) & 0.363 (0.032) & 0.692 (0.093) & 0.475 (0.050) \\
		\hline
		& 100  & 1.524 (0.315) & 0.954 (0.156) & 2.547 (0.571) & 1.695 (0.362) \\
		$0.1-0.5$     & 300  & 0.933 (0.151) & 0.712 (0.097) & 1.925 (0.424) & 1.271 (0.243) \\
		& 3000 & 0.412 (0.039) & 0.397 (0.037) & 1.080 (0.190) & 0.710 (0.097) \\
		\hline
	\end{tabular}
	\caption{The $95\protect\%$ C.L. constraints on the anomalous $HZ\gamma$ couplings, $|\alpha_{1}|$ and $|\alpha_{2}|$, and the upper limits on the branching ratio of the $H\rightarrow Z\gamma$ decay channel in the main process $pp\rightarrow pHX$ at $\sqrt{s}=14$ TeV and for integrated luminosities $\mathcal{L}_{\mbox{\tiny{int}}}=100,300,3000$ fb$^{-1}$. The upper limits on the branching ratio are given in parentheses. The bounds values are presented for three different Higgs decay channels, $H\rightarrow\gamma\gamma$, $H\rightarrow W^{+}W^{-}$, and $H\rightarrow ZZ$, as well as a combined one achievable with $95\protect\%$ C.L. and for three intervals of forward detector acceptance region, $\xi$. The CP-even and CP-odd contributions have the same values. All the reconstruction and the acceptance efficiencies as well as pile-up backgrounds are included into bounds estimations.}
	\label{tab4}
\end{table}
\begin{table}[h]
	\begin{tabular}{|c||c|c|c|c|c|c|c|c|c|c|}\hline
		
		&  & \multicolumn{4}{c|}{$|\alpha_{1}|=|\alpha_{2}|$\ [TeV$^{-1}$]\ \ $\big(Br(H\to\gamma Z)\big)$} \\\hline
		$\xi$ & \backslashbox{\ \ \ $\mathcal{L}_{\mbox{\tiny{int}}}$[fb$^{-1}$]}{Channel} &$H\rightarrow \gamma\gamma$ &$H\rightarrow ZZ$ & $H\rightarrow W^{+}W^{-}$ & Combined \\ \hline\hline
		
		& 100  & 0.389 (0.036) & 0.279 (0.021) & 0.671 (0.088) & 0.446 (0.045) \\
		$0.0015-0.5$  & 300  & 0.256 (0.019) & 0.210 (0.014) & 0.508 (0.055) & 0.336 (0.028) \\
		& 3000 & 0.129 (0.0075) & 0.117 (0.0067) & 0.285 (0.022) & 0.188 (0.012) \\
		\hline
		& 100  & 0.414 (0.040) & 0.292 (0.023) & 0.683 (0.091) & 0.455 (0.046) \\
		$0.0015-0.15$ & 300  & 0.274 (0.021) & 0.219 (0.015) & 0.517 (0.057) & 0.342 (0.029) \\
		& 3000 & 0.139 (0.008) & 0.123 (0.007) & 0.290 (0.023) & 0.192 (0.012) \\
		\hline
		& 100  & 0.792 (0.116) & 0.330 (0.028) & 1.021 (0.174) & 0.695 (0.093) \\
		$0.1-0.5$     & 300  & 0.457 (0.047) & 0.212 (0.014) & 0.761 (0.108) & 0.506 (0.055) \\
		& 3000 & 0.145 (0.009) & 0.104 (0.006) & 0.424 (0.041) & 0.278 (0.021) \\
		\hline
	\end{tabular}
	\caption{The $95\protect\%$ C.L. constraints on the anomalous $HZ\gamma$ couplings, $|\alpha_{1}|$ and $|\alpha_{2}|$, and the upper limits on the branching ratio of the $H\rightarrow Z\gamma$ decay channel in the main process $pp\rightarrow pHX$ at $\sqrt{s}=100$ TeV and for integrated luminosities $\mathcal{L}_{\mbox{\tiny{int}}}=100,300,3000$ fb$^{-1}$. The upper limits on the branching ratio are given in parentheses. See the caption of Table \ref{tab4} for further details.}
	\label{tab5}
\end{table}
\par
\begin{figure}[ht]
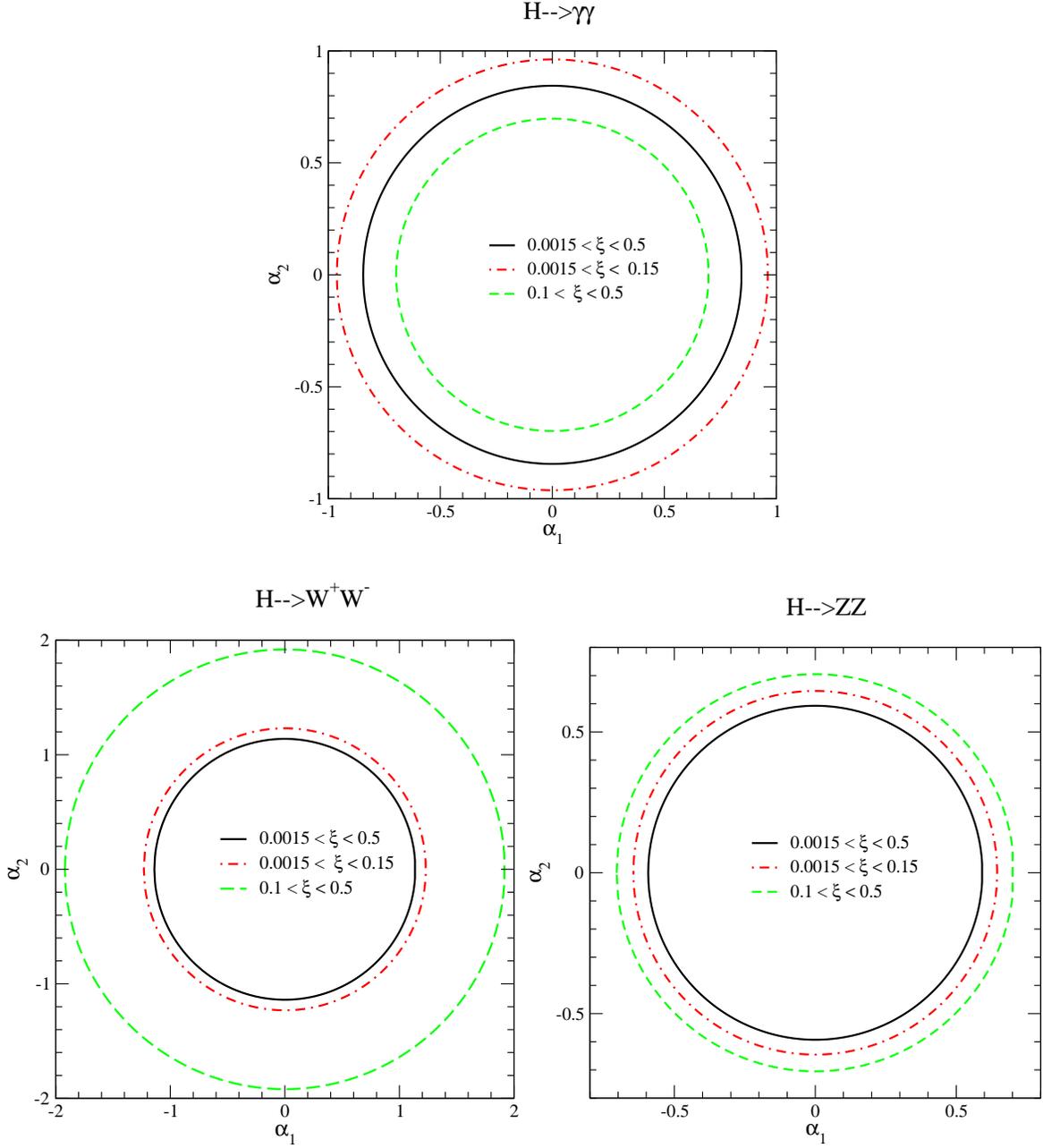

\centerline{
\includegraphics[clip,width=0.45\textwidth]{gammagamma.eps}}
\par
\vspace{0.5cm}
\centerline{
\includegraphics[clip,width=0.45\textwidth]{ww.eps}
\includegraphics[clip,width=0.45\textwidth]{zz.eps}}
\caption{The contour diagrams in $\alpha_2-\alpha_1$ plane (units in TeV$^{-1}$) for three different Higgs decay channels $H\rightarrow\gamma\gamma$, $H\rightarrow W^{+}W^{-}$, and $H\rightarrow ZZ$ with $95\protect\%$ C.L. at $\sqrt{s}=14$ TeV and $\mathcal{L}_{\mbox{\tiny{int}}}=300$ fb$^{-1}$. The diagrams are plotted for three different acceptance regions. Both the reconstruction and the acceptance efficiencies are included into bounds estimations.}
\label{Constraint}
\end{figure}

Fig. \ref{Constraint} illustrates the contour diagrams for the $95\%$ C.L. constraints on the anomalous couplings in the $\alpha_{2}$--$ \alpha_{1}$ plane for three different Higgs decay channels $H\rightarrow \gamma \gamma $, $H\rightarrow W^{+}W^{-}$, and $H\rightarrow ZZ$ at $\sqrt{s}=14~\mbox{TeV}$ and $\mathcal{{L}}_{\text{{int}}}= 300~\mbox{fb}^{-1}$. The diagrams are plotted for three different acceptance regions while both the reconstruction and the acceptance efficiencies as well as pile-up backgrounds are included. Each panel contains the results of a specific Higgs decay channel.
\par
A similar analysis on search for the anomalous $HZ\gamma $ couplings, which only concentrates on the $H\rightarrow b\bar{b}$ channel has been performed in Ref.~\cite{Senol:2014naa}. In that study, ignoring the irreducible backgrounds, the authors have only considered the reducible ones, so their analysis has consequently lead to tight bounds at the level of $\sim 10^{-3}$. In this paper, by taking into account the most relevant backgrounds (the reducible part) as well as the irreducible ones, and looking at the clean decay modes, i.e., $\gamma \gamma $, $W^{+}W^{-}$, and $ZZ$, more realistic results are obtained. The present analysis of the process $pp\rightarrow pHX$ could be potentiality considered as a first assessment of the LHC to study the $HZ\gamma $ couplings.
\par
Finally, it is necessary to emphasize that in this paper a simple counting experiment analysis has been performed to obtain the upper limits on the anomalous couplings and branching ratios. It is notable that in some cases, the generality of the couplings affects the kinematic distributions of the final state particles. Therefore, the kinematic distributions provide powerful discriminating variables among various anomalous couplings of signal and background processes. Similar to the ATLAS and CMS experiments, following smart methods such as matrix element likelihood approach would provide more stringent bounds. The mentioned approach is useful to construct a discriminant for the analysis of the kinematic distributions of the Higgs boson production and decay in different channels \cite{cmszz}. However, this is beyond the scope of the present paper and must be done by the experimental collaborations to include detailed simulation effects and detector response.

\section{Concluding remarks}\label{sec5}
After the discovery of the SM Higgs boson at the LHC, direct and indirect searches are ongoing for precise measurements of the Higgs boson properties. The purpose of this paper is to examine the potential of the Higgs boson photoproduction at the LHC to probe the anomalous $HZ\gamma $ couplings originating from dimension six non-SM operators. We study the deviations of both CP-even and CP-odd anomalous $HZ\gamma $ couplings from the SM predictions, which arise from NP effects. To this end we established precise bounds on the anomalous couplings for three different detector acceptance regions, $0.0015<\xi <0.5$, $0.0015<\xi <0.15$, and $0.1<\xi <0.5$. We have predicted that the future LHC run has a good capability to establish the CP nature of the $HZ\gamma $ vertices using the detectors that would be available in the forward regions. The total cross section of the studied process $pp\rightarrow pHX$ shows similar sensitivity to the CP-even and CP-odd couplings. Since the angular distributions of the decay products of the Higgs boson have different behaviors for the CP-even and CP-odd couplings, they could be used as powerful tools to examine the CP nature of the couplings. Here using a simple counting experiment analysis, the first and second acceptance regions, i.e., $0.0015<\xi <0.5$ and $0.0015<\xi <0.15$, provide the most restricted bounds in combined channel. At the LHC, with an integrated luminosity $\mathcal{L}_{\mbox{{int}}}=3000~\mbox{fb}^{-1}$ at $\sqrt{s}=14~\mbox{TeV}$ while both the reconstruction and the acceptance efficiencies are included, the bounds on anomalous $HZ\gamma $ vertices for the first region would be 0.475, 0.333, and 0.640 in $H\rightarrow \gamma \gamma $, $H\rightarrow ZZ$, and $H\rightarrow W^{+}W^{-}$ decay channels, respectively. The best limits on $HZ\gamma $ couplings are obtained from $H\rightarrow ZZ$ channel. We conclude that the process $pp\rightarrow pHX$ has a reasonable sensitivity to the anomalous $HZ\gamma $ couplings which complements the results of other channels in search for any deviation of $HZ\gamma $ vertices from the SM predictions.

\section*{ACKNOWLEDGMENTS}\label{sec6}
The authors are thankful to the School of Particles and Accelerators, Institute for Research in Fundamental Sciences (IPM). S. T. M. and Sh. F. gratefully acknowledge partial support of this research provided by the Islamic Azad University Central Tehran Branch. 


\end{document}